\documentclass[]{aa} 

\usepackage{graphicx}
\usepackage{txfonts}

\begin{document}
\title{Emission lines from rotating proto-stellar jets with variable 
velocity profiles}

\subtitle{I. Three-dimensional numerical simulation of the non-magnetic case}

\author{
A.H. Cerqueira \inst{1,~2}, 
P.F. Vel\'azquez \inst{2}, 
A.C. Raga \inst{2}, 
M.J. Vasconcelos \inst{1,~2},
\and \\
F. De Colle \inst{3}
       }

\offprints{A.H. Cerqueira}

\titlerunning{Emission lines from rotating jets}
\authorrunning{A. H. Cerqueira et al.}

\institute{
LATO-DCET-UESC, Rodovia Ilh\'eus-Itabuna km 16, Ilh\'eus, Bahia, 45662-000, Brazil\\
\email{hoth@uesc.br, mjvasc@uesc.br}
\and
Instituto de Ciencias Nucleares, UNAM, Ap. Postal 70-543, CU, D.F., 04510, M\'exico\\
\email{pablo@nucleares.unam.mx, raga@nucleares.unam.mx}
\and
Instituto de Astronom{\'{\i}}a, UNAM, Ap. Postal 70-543, CU, D.F., 04510, M\'exico\\
\email{fdecolle@astroscu.unam.mx}
          }

\date{Received August 26, 2005; accepted October 17, 2005}

\abstract{Using the Yguaz\'u-a three-dimensional hydrodynamic code, we
have computed a set of numerical simulations of heavy, supersonic,
radiatively cooling jets including variabilities in both the
ejection direction (precession) and the jet velocity (intermittence). In
order to investigate the effects of jet rotation on the shape of the
line profiles, we also introduce an initial toroidal rotation velocity profile,
in agreement with some recent observational evidence found in jets from
T Tauri stars which seems to support the presence of a rotation velocity
pattern inside the jet beam, near the jet production region. Since
the Yguaz\'u-a code includes an atomic/ionic network, we are able
to compute the emission coefficients for several emission lines,
and we generate line profiles for the H$\alpha$, [O I]$\lambda$6300,
[S II]$\lambda$6716 and [N II]$\lambda$6548 lines.  Using initial
parameters that are suitable for the DG Tau microjet, we show that the
computed radial velocity shift for the medium-velocity component of the
line profile as a function of distance from the jet axis is strikingly
similar for rotating and non-rotating jet models.  These findings lead us
to put forward some caveats on the interpretation of the observed radial
velocity distribution from a few outflows from young stellar objects,
and we claim that these data should not be directly used as a doubtless
confirmation of the magnetocentrifugal wind acceleration models.

            }

\authorrunning{Cerqueira et al}
\maketitle
%

\section{Introduction}

It is now widely accepted that Herbig-Haro jets are a common by-product of
the formation of a low mass star. Initially, it was proposed that these
jets originate as stellar winds (\cite{canto80,hart..82}). However, the
apparent correlation between ejection and accretion signatures favoured
disk wind models. Blandford \& Payne (\cite{bland82}) proposed a model in
which a wind is launched from the disk surface through open magnetic field
lines which make an angle greater than $30^{\circ}$ with the vertical
direction. This was called the magneto-centrifugal model because the
centrifugal force exerted in a particle attached to an open magnetic field
line (satisfying the condition mentioned above) accelerates it to speeds
above the escape velocity.  More recently, Ferreira (\cite{ferreira97})
and Casse \& Ferreira (\cite{casse..00a}; \cite{casse..00b}) proposed
self-similar jet launching mechanisms arising from an extended region of
the accretion disk in which the magnetic field lines have their origin in
the disk itself.  They analyzed {\it cold} and {\it warm} models. These
models are ruled by three dimensionless parameters: the disk aspect
ratio $\varepsilon = h/r$, where $h$ is the disk scale height at the
cylindrical radius $r$, the turbulence level $\alpha_m = \nu_m / V_A h$
(where $\nu_m$ is the turbulent magnetic diffusivity and $V_A$ is the
Alfv\'en velocity at the disk midplane), and the ejection index $\xi =$
d ln $\dot{M}_a$ / d ln $r$ (where $\dot{M}_a$ is the accretion rate),
which gives the efficiency of the launching process. In the cold models,
the enthalpy at the jet basis is null, while the warm jet launching
models have some heating source that provides additional thermal energy.
The ejection efficiency for cold jets is at most $\xi \sim 0.01$ while
warm jets could have efficiencies of up to $\xi \sim 0.5$. Another kind
of disk model is the so called {\it X-wind} (Shu et al. \cite{shu..94}).
In this case, the disk is truncated exactly at the corotation point
and the jet is launched from this truncation radius. In both models,
the rotation of the disk plays a very important role. The exact point
or region from which the jet arises is still under debate.

Although the magneto-centrifugal acceleration (hereafter, MCA) is
widely accepted as being the general mechanism behind the launching
of a disk wind, the lack of observational support for it
until recently has contributed to keep the subject open. In the last
few years, however, observations of a few jets associated with young T
Tauri stars appear to support this model, since the interpretation
of the radial velocity profiles close to the launching region is
both qualitatively and quantitatively in agreement with them (e.g.,
Bacciotti et al. \cite{bacci02}; Coffey et al. \cite{coffey04}; Woitas et
al. \cite{woitas05}). These authors claim that, for a few jets associated
with T Tauri stars (namely DG Tau, Th28, RW Aur and LkH$\alpha$321),
there is a trend in the radial velocity pattern which indicates that
the jet is rotating. Furthermore, they find that the rotation velocity
implied by the observed radial velocity distribution is in good agreement
with the rotational profile expected for a wind launched by the MCA model
(see Pesenti et al. \cite{pesenti}).

Bacciotti et al. (\cite{bacci02}) have carried out HST/STIS measurements
of the [O I]$\lambda\lambda$6300,6363 and [S II]$\lambda\lambda$6716,6731
emission line profiles of the DG Tau jet. The observation was designed
to construct a datacube, changing the position of the long slit (of
$0\farcs1\times52\farcs$) several times.  The datacube was then used to
determine the spatial distribution of the radial velocity. In particular,
they have taken spectra for seven different positions perpendicular to
the DG Tau jet, for a given distance from the source, by offsetting each
slit position by $0\farcs07$. Assuming that the jet is axisymmetric, they
carefully determine the position of the jet axis, by determining the peak
of each emission line, and radial velocities associated with symmetrically
disposed positions (on both sides of the jet axis) were measured. The line
profiles obtained by Bacciotti et al. (\cite{bacci02}) were decomposed
into two components, and the medium-velocity component (hereafter, MVC) was
used to perform the analysis.  The radial velocity (associated with the
MVC) was then plotted against the slit position (i.e., distance from the
jet axis), for four regions of increasing distance from the source (named
regions I, II, III and IV, at distances from the source of $0\farcs075$,
$0\farcs175$, $0\farcs275$ and $0\farcs4$, respectively). They found
that the radial velocity shift for positions symmetrically distant from
the jet axis is systematically negative, regardless of the chosen region
(I, II, III or IV) or the emission line\footnote{There are, in fact,
some points in their data that do not follow such a trend, but this might
be due to either small fluxes or the proximity to the bow shock of the
DG Tau jet; see the discussion in Bacciotti et al. (\cite{bacci02}).}.
This systematic negative shift in the radial velocity differences was
interpreted by the authors as evidence that the jet is rotating.

The values found for the rotational velocity of the DG Tau jet, in the
range of 6-15 km s$^{-1}$, are in agreement with the values predicted by
the MCA models, for an assumed $\dot{M}_{jet}/\dot{M}_{acc} \sim 0.1$
(which is a bona fide value for these young stellar systems and,
in particular, for the DG Tau system).  This result, if combined
with the fact that the circunstelar disk that surrounds DG Tau shows
a velocity pattern [extracted from the $^{13}$CO(2-1) data by Testi
et al. \cite{testi}] that is consistent with a disk rotating in the
same direction as the jet, reinforces the MCA scenario. Applying a
technique very similar to that of Bacciotti et al. (\cite{bacci02}),
Woitas et al. (\cite{woitas05}) have re-analyzed the data from their
HST/STIS observations of the RW Aur bipolar flow (see Woitas et
al. \cite{woitas02}), and they found the same trend in the radial
velocity shifts.

Coffey et al. (\cite{coffey04}), also using the HST/STIS spectrograph,
found that the bipolar jets from the T Tauri stars TH28 and RW Aur (see
also Woitas et al. \cite{woitas05} for the latter source), as well as
the blue-shifted jet from LkH$\alpha$321 also show systematic velocity
asymmetries which are consistent with the results obtained by Bacciotti
et al. (\cite{bacci02}) for DG Tau. In other words, they also interpret
the radial negative (or positive) velocity shifts as a signature of
rotation (clockwise or counter-clockwise, depending on the sign of the
radial velocity shift; see Coffey et al. \cite{coffey04} for details).
In particular, they found that in the TH28 and RW Aur flows, which
are bipolar, both the jet and the counter-jet appear to rotate in the
same direction, and they use this fact as an argument in favour of the
MCA models.

These observational data and, more importantly, their interpretation as
a jet rotation signature, lead some authors to use them to constrain the
wind launching region on the surface of the accretion disk. Pesenti et
al. (\cite{pesenti}) used their self-similar MHD disk wind solutions to
investigate the possible influence of projection or excitation gradients
in the relation between the rotation signature and the true azimuthal
velocity profile. They computed cold and warm models with different
parameters and found that warm models, with an outer radius\footnote{This
is the outer radius of the disk region responsible for the launching of
the wind.} $\sim$ 3 AU, better reproduce the observations of DG Tau. They
found that both cold and warm disk solutions predicts increasing values
for the radial velocity shifts for increasing distances from the jet
axis. However, the values predicted for the velocity shifts in the warm
solution are systematically smaller than those predicted by the cold
disk solutions, in better agreement with the observations. They also
argue that the shift in the radial velocity for the computed models
are systematically smaller than the values given by the true rotational
profiles, and that this effect is more pronounced near the jet axis.

In a somewhat different approach, Cerqueira \& de Gouveia Dal Pino
(\cite{cerq04}) have performed three-dimensional SPH numerical
simulations of rotating jets. They were able to reproduce the trend in
the shift of the radial velocities observed in DG Tau by Bacciotti et
al. (\cite{bacci02}), with a model of a precessing (half-opening angle
of $5^{\circ}$), variable velocity (300 $\pm$ 100 km s$^{-1}$, sinusoidal
velocity variation with a $\tau_{pul} = 8$ years period) jet. Cerqueira \&
de Gouveia Dal Pino (\cite{cerq04}) have not computed the emission line
profiles, that would allow a direct comparison between the numerical
models and the observed data. These authors claim that a magnetic field
of intensity of the order of $\approx 1$ $\mu$G is necessary in order to
maintain jet stability against lateral expansion due to centrifugal
forces. Actually, very recent modeling of observational data carried out
by Woitas et al. (\cite{woitas05}) for both lobes of the bipolar
outflow from RW Aur also suggest a magnetic collimation mechanism
for the jet.

These observations and models are increasingly providing support for a
scenario in which the MCA model matches the available observational data
(considering the errors estimated from both).  However, apart from the
very low number of observed sources with this kind of radial velocity
pattern (four until now), there are still a few points that should be
carefully verified before excluding alternative views, which can also
explain the observed data. For example, Soker (\cite{soker}) proposed that
the interaction between the jet with a warped accretion disk can explain
the asymmetry in the radial velocity distribution across the jet beam,
although his model is in need of direct observational support (since an
inclination between the jet and the outer parts of the accretion disk
is assumed).

In this paper, we present the results of a set of fully three-dimensional
numerical simulations of a non-magnetic\footnote{The discussion of
the magnetized case will be left to a forthcoming paper (De Colle
et al. 2005; in preparation).} jet that is allowed to precess and
rotate. This, together with an ejection velocity time-variability (which
produces multiple internal working surfaces), represents a complete
scenario for the Herbig-Haro jets, which are believed to present the
observational signature of rotation. With a proper calculation of the
emission coefficients, the cooling of the gas, and the evolution of the
atomic/ionic species, we are able to produce velocity channel maps,
and then reconstruct the line profiles for several of the observed
forbidden and permitted (namely H$\alpha$) emission lines. We then make
a comparison between the radial velocities extracted from these computed
line profiles and those from previously published data of HH jets.

The paper is organized as follows. In \S 2, we describe the numerical
technique and the physical parameters of the simulations.  The results
are presented in \S 3, and in \S 4 we discuss these results and their
relation with the available observational data.

\section{The numerical method and the simulated models}

\subsection{The Yguaz\'u-a code}

The 3-D numerical simulations have been carried out with the
Yguaz\'u-a adaptive grid code (which is described in detail by Raga et
al. \cite{raga00} and Raga et al. \cite{raga02}), using a 5-level binary
adaptive grid.  The Yguaz\'u-a code integrates the gas-dynamic equations
(employing the ``flux vector splitting'' scheme of van Leer \cite{van})
together with a system of rate equations for the atomic/ionic species:
HI, HII, HeI, HeII, HeIII, CII, CIII, CIV, NI, NII, NIII, OI, OII, OIII,
OIV, SII and SIII.  With these rate equations, a non-equilibrium cooling
function is computed. The reaction and cooling rates are described in
detail by Raga et al. (\cite{raga02}).

This code has been extensively employed for simulating several
astrophysical scenarios such as jets (Masciadri et al. \cite{mas};
Vel\'azquez, Riera \& Raga \cite{vela1}; Raga et al. \cite{raga04};
Raga et al. \cite{raga02}), interacting winds (Gonz\'alez et al
\cite{gon04a,gon04b}; Raga et al. \cite{raga01a}) and supernova remnants
(Vel\'azquez et \cite{vel04}; Vel\'azquez et al. \cite{vel01a}). It
was also tested with laser generated plasma laboratory experiments (Raga et
al. \cite{raga01b}; Sobral et al.  \cite{sobral}; Vel\'azquez et
al. \cite{vel01b}).

The computational domain has dimensions of $1.1\times10^{16}$ cm along
the $x-$ and $y-$axes, and of $4.4\times10^{16}$ cm along the $z-$axis.
The grid has a maximum resolution of $8.59\times10^{13}$ cm along the
three axes. Assuming a distance of 140 pc (e.g., \cite{Kenyon}), this
maximum resolution corresponds to an angular size of $\approx 0\farcs04$.

\subsection{The models}

\begin{table}
\caption[]{The simulated models.}
\label{tab1}
$$ 
\begin{array}{p{0.2\linewidth}cccc}
\hline
Model &  \tau_{pul.} & \tau_{pre.} & \theta & {\rm rotation?} \\
\hline
M1    &       8      &       -     &    -   &     {\rm no}    \\
M2    &       8      &       8     &    5   &     {\rm no}    \\
M3    &       8      &       -     &    -   &     {\rm yes}   \\
M4    &       8      &       8     &    5   &     {\rm yes}   \\
\hline
\end{array}
$$ 
\end{table}

The parameters were chosen in order to model the observations of the DG
Tau jet. All the simulated models have a time-dependent ejection velocity
given by (Raga et al. \cite{raga01c}):

\begin{equation}\label{vjet}
v_j=v_0\left[1+A~{\rm sin}\left(\frac{2\pi}{\tau_{pul}}t\right)\right],
\end{equation}

\noindent where $v_0=300$ km s$^{-1}$, and $A=0.33$ (giving a jet velocity
variability in the range of 200-400 km s$^{-1}$) and $\tau_{pul} = 8$
years. These parameters are appropriate for producing the velocity
structure of the DG Tau microjet (Bacciotti et al. \cite{bacci02};
Raga et al. \cite{raga01c}; Pyo et al. \cite{pyo}). There is also
evidence that the DG Tau microjet precesses, with a precessing angle of
$\approx 5^{\circ}$ and a precessing period of $\approx 8$ years (e.g.;
Lavalley-Fouquet et al. \cite{lava2}; Dougados et al. \cite{dou00}).
In Table \ref{tab1} we give the model label (first column), the period
(in years) for the velocity variability and for the precession of the
jet axis (second and third column, respectively), the half-opening
angle of the precession cone (fourth column) and, finally, the presence
(or not) of rotation in a given model, expressed by a toroidal velocity
profile (see below). The physical parameters of the models presented in
Table \ref{tab1} are, thus, suitable for the study of the DG Tau jet.
Finally, we choose an initial number density for the jet $n_j=1000$
cm$^{-3}$ (with $\eta=n_j/n_e=10$, where $n_e$ is the initial number
density for the environment), and a temperature $T_j=10^4$ K (with
$\kappa=T_j/T_e=10$). We also assume that the initial jet H ionization
fraction is of 10\%.

In order to investigate the signatures of rotation on the line profiles,
we have simulated models with and without an initially imposed toroidal
velocity profile (see Table \ref{tab1}). The adopted toroidal velocity
is the same as in Cerqueira \& de Gouveia Dal Pino (\cite{cerq04}):

\begin{equation}\label{vphi}
v_{\phi} = 8 {\rm km ~ s}^{-1} \cdot \frac{R_j}{R},
\end{equation}

\noindent where $v_{\phi}$ is the toroidal velocity in km s$^{-1}$,
$R_j$ is the jet radius ($5.6\times10^{14}$ cm in physical units) and $R$
is the cylindrical radius. We note that this profile has been truncated
at a radial distance $= 0.15R_j$, where the toroidal velocity attains
a constant value of $v_{\phi}\simeq 55$ km s$^{-1}$.

We should note that the radial dependence of the toroidal velocity
depends on the details of the MCA wind model that is considered. For
example, the model of Pesenti et al. (2004) predicts a $v_{\phi}\propto
R^{-1/2}$ law. In the present paper, we restrict ourselves to the toroidal
velocity law given by equation (\ref{vphi}), and do not explore other
possibilities.

From the temperature, and atomic/ionic/electronic number densities
computed in the numerical time-integration, we calculate the emission
line coefficients of a set of permitted and forbidden emission lines.
We compute the Balmer H$\alpha$ line considering the contributions from
the recombination cascade and from $n=1\to 3$ collisional excitations. The
forbidden lines are all computed by solving 5-level atom problems,
using the parameters of Mendoza (\cite{mendoza}).

\section{Results}

\begin{figure*}
\centering \hskip -3truecm 
\includegraphics[width=15cm]{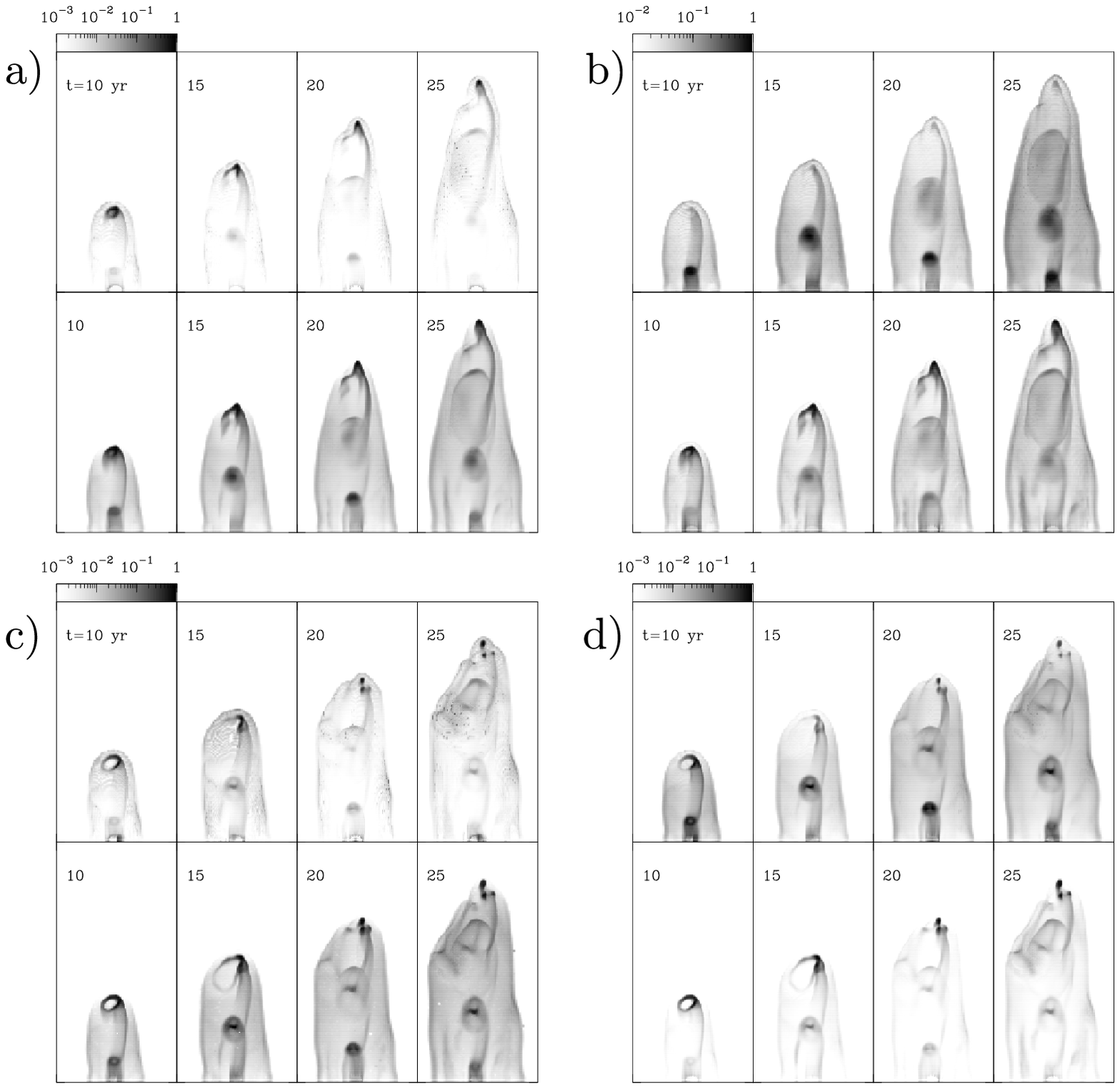}
\vskip -10truecm
\caption{
{\bf a)} Temporal evolution ($t=10$, 15, 20 and 25 years, from
left to right) of the emission maps for the H$\alpha$ (top) and [S
II]$\lambda$6716 (bottom) lines, computed for model M2 (see Table
\ref{tab1}).
{\bf b)} The same as in $a)$ for the [O I]$\lambda$6300 (top) and
[N II]$\lambda$6548 (bottom) lines.
{\bf c)} Temporal evolution ($t=10$, 15, 20 and 25 years, from
left to right) of the emission maps for the H$\alpha$ (top) and [S
II]$\lambda$6716 (bottom) lines, computed for model M4 (see Table
\ref{tab1}).
{\bf d)} The same as in $c)$ for the [O I]$\lambda$6300 (top) and [N
II]$\lambda$6548 (bottom) lines. The jet in each map was assumed to have
an angle with respect to the line of sight of $45^{\circ}$. The bars
on top of each diagram give the normalized flux.
[See the electronic edition of Astronomy \& Astrophysics for a color
version of this figure.]
         }

\label{emaps}
\end{figure*}

Figure \ref{emaps} shows the temporal evolution for the emission maps of
models M2 (Fig \ref{emaps}a, b) and M4 (Fig \ref{emaps}c, d), assuming
an inclination angle with respect to the line of sight of 45$^{\circ}$
(which is the estimated inclination angle of DG Tau; see Pyo et
al. \cite{pyo}).  Model M2 presents a variability in both direction
and ejection velocity, and model M4 also has an initially imposed
toroidal velocity pattern (see \S 2). The computed emission lines are:
H$\alpha$ and [S II]$\lambda$6716 (left panels), [O I]$\lambda$6300 and
[N II]$\lambda$6548 (right panels). The small precession angle (see Table
\ref{tab1}) in both models produces a gentle wiggle along the jet beam,
an effect that can be easily seen in Figure \ref{emaps}. The internal
working surfaces are morphologically similar in both models.

The major difference between the M2 and M4 models lies in the morphology
of the jet head. In the rotating jet model M4, the jet head appears to be
much less collimated, an effect that is due to the centrifugal force that
increases the jet opening angle.  This effect is clearly seen through a
direct comparison between the emission maps of models M4 and M2 in the
head region. For model M2, Figures \ref{emaps}a and b shows a collimated
jet head, while the jet head of model M4 in Figures \ref{emaps}c and d
shows a clumpy, fragmented structure. Furthermore, there is a little
deceleration of the jet head in model M4 with respect to model M2,
an effect that is associated with the de-collimation of the jet head.

For each evolutionary time, and for all of the computed emission lines
(namely, H$\alpha$, [S II]$\lambda$6716, [O I]$\lambda$6300 and [N
II]$\lambda$6548), we compute velocity channel maps in the -400 to 100
km s$^{-1}$ range (with a $resolution$ of 10 km s$^{-1}$), assuming an
inclination angle of $45^{\circ}$ towards the observer. These velocity
channel maps allow us to reconstruct the line profiles, since we have
the intensity of the line for a given radial velocity as a function of
position in the plane of sky (see \S 2). Thus, we use this $datacube$
to build line profiles for selected regions in our computational
domain. In particular, we use $3\times3$ px$^{2}$ region to define a
single, $artificial~slit$ (which, for an assumed distance of 140 pc
and an assumed inclination angle with respect to the line of sight of
45$^{\circ}$, gives us a $\approx 0\farcs1\times0\farcs1$ slit). An
intensity is determined for each velocity channel, as a function of
the slit position in the computational domain. These intensity points,
at 10 km s$^{-1}$ radial velocity intervals, are then combined to give
the line profile for each slit (as a function of position).

\begin{figure*}
\centering \hskip -1.2truecm
\includegraphics[width=14cm]{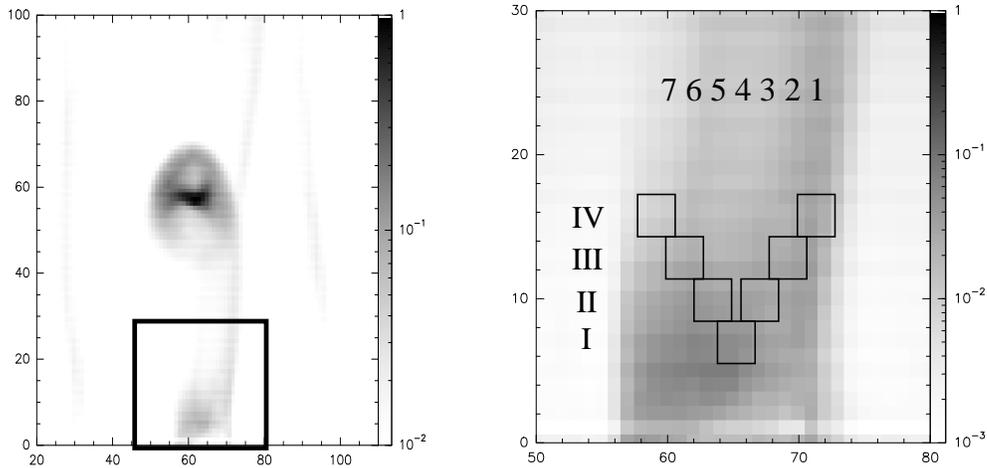}
\caption{{\bf left)} An [O II]$\lambda$6300 line emission map for the M4
model in a region near to the jet inlet, at t=15 years.  The coordinates
are in units of pixels, and the scale bar indicates the normalized
intensity. {\bf right)} A closer view of the jet base, corresponding
to the region indicated by the rectangular box in the left-panel,
indicating also the $artificial$ slit positions. Regions I, II, III
and IV are successively more distant from the jet inlet, while positions
1, 2, 3, 4, 5, 6 and 7 corresponds to different position across the jet
radius. Note that each slit have a superposition of 1 px in the direction
transverse to the jet axis. For the sake of clarity, we show in this
figure only a few (one or two) of the seven slits for each region. See
the text for discussion.
[See the electronic edition of Astronomy \& Astrophysics for a color
version of this figure.]
         }
\label{slits}
\end{figure*}

We define four regions along the jet axis, namely regions I, II, III
and IV, at distances from the jet inlet of $0\farcs1$, $0\farcs2$,
$0\farcs3$ and $0\farcs4$, respectively (assuming a distance of
140 pc and an inclination angle of 45$^{\circ}$;
see \S 2). For each one of these regions, we construct seven
$0\farcs1\times0\farcs1$ slits (as described above), that are placed
side by side parallel to the jet axis with one of them lying exactly
on the jet axis (in order to have slits symmetrically disposed on
each side of the jet). This configuration (seven aligned slits across
the jet beam and in four regions along the jet beam) was chosen in
order to mimic the observational conditions described in Bacciotti et
al. (\cite{bacci02}). In Figure \ref{slits} we show the slits and their
positions on the plane of the sky. We note that, in fact, there is an
overlap of 1 pixel ($\approx 0\farcs07$) between the slits in each region
(see Figure \ref{slits}).

\begin{figure*}
\centering
\hskip -1truecm
\includegraphics[width=14cm]{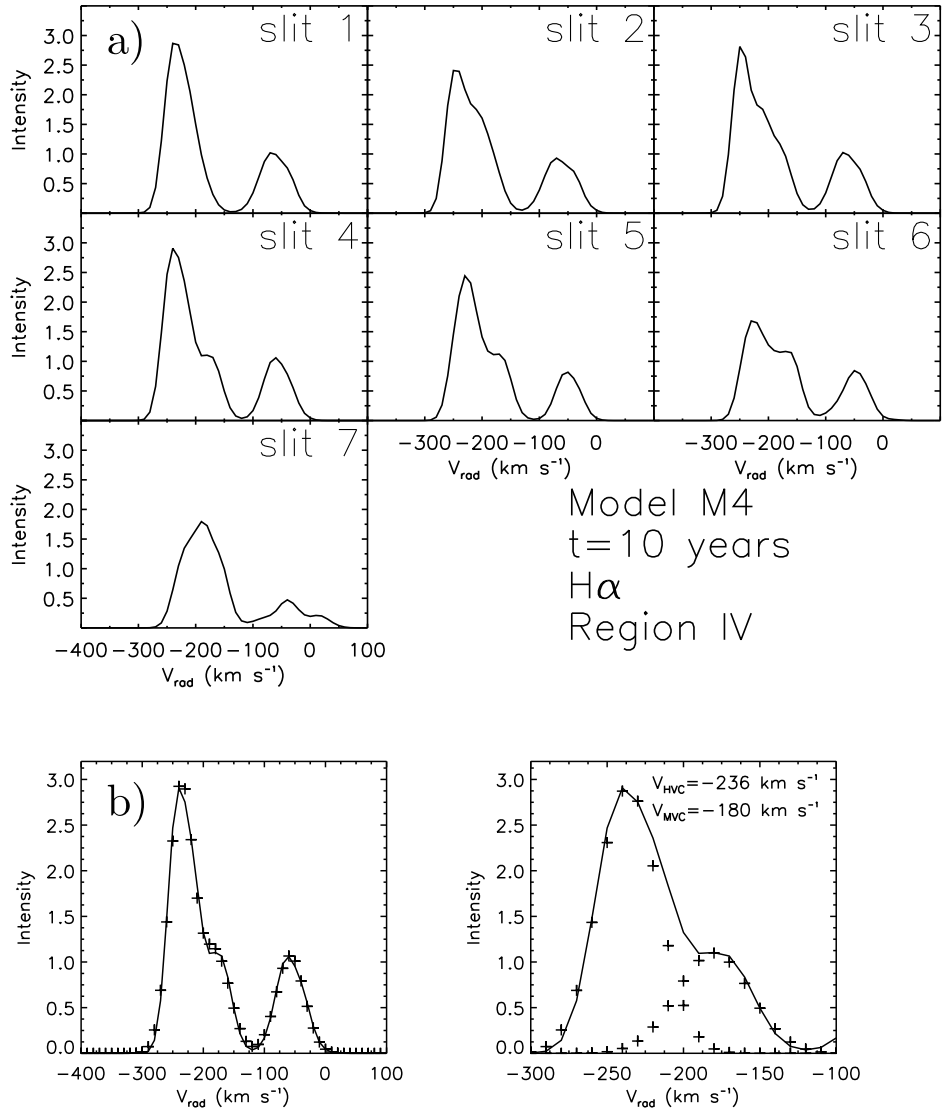}
\vskip -5truecm
\caption{
{\bf a)} H$\alpha$ line profile for model M4, at $t=10$ years, for region
4 (see Figure \ref{slits} for slit and region definitions). The slit
number is given on the top-right corner of each diagram. The intensity
is in units of $10^{-6}$ erg cm$^{-2}$ s$^{-1}$.  {\bf b)} {\it Left:}
The H$\alpha$ line profile of slit 4, region IV (full line), superimposed
with a three-component $\chi^2$-gaussian fit (crosses).  {\it Right:}
a closer viewer of the line profile in the range of -300 to -100 km
s$^{-1}$, showing the two gaussians (crosses) defined as the high-
and medium-velocity components (HVC and MVC, respectively).
See the text for a discussion.
         }
\label{prof}
\end{figure*}

In Figure \ref{prof}a, we show the H$\alpha$ line profiles for region
IV of model M4 at $t=10$ yr. We note the presence of at least three
components in several of the line profiles shown in Figure \ref{prof}. The
emission in the radial velocity range from $\approx$ 0 to $\approx$ -100
km s$^{-1}$ is related to the leading bow shock, which surrounds the jet
beam. We assume that the emission in the interval from -300 km s$^{-1}$
to -100 km s$^{-1}$ is related to two components, namely, the high-
and medium-velocity components (hereafter, HVC and MVC, respectively). We
have performed standard minimum $\chi^2$-Gaussian fits to obtain the
MVC, for the seven slits in each of the four regions defined in Figure
\ref{slits}. Figure \ref{prof}b shows the data (full line) combined
with the gaussian fit (crosses) for the whole radial velocity interval
(left) and for the radial velocity interval ranging from -300 km s$^{-1}$
to -100 km s$^{-1}$ (right). We also show the gaussian fits decomposing
the HVC and MVC. We have performed these fits for all models in Table
\ref{tab1}, and for different evolutionary times.  This procedure allows
us to investigate both the behaviour of the MVC for each emission line
as well as its dependence with the imposed initial conditions.

\begin{figure*}
\centering \hskip -3truecm
\includegraphics[width=16cm]{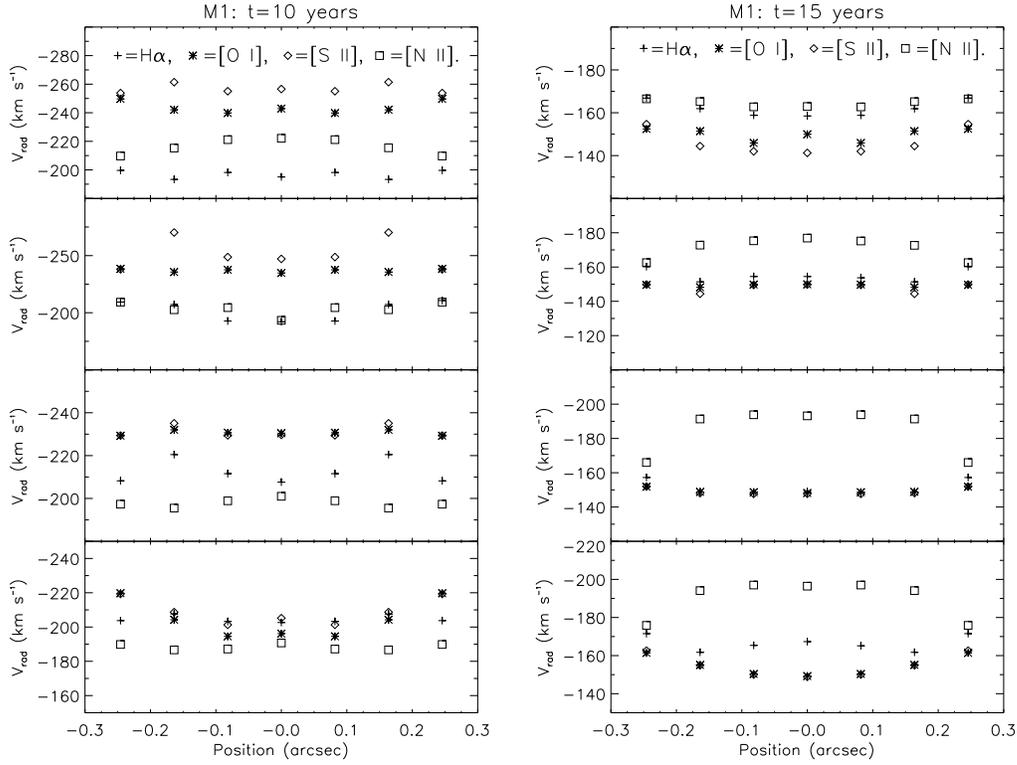}
\vskip -12truecm
\caption{Radial velocity as a function of distance from the jet axis
for regions I to IV (from top to bottom), for $t=10$ years (left) and
$t=15$ years (right) and model M1 (see Table \ref{tab1}). The distance is
in units of arcsec, considering a distance of 140 pc from the jet (the zero
corresponds to the jet axis position, and the scale, ranging from negative
to positive values, corresponds to slits 1 to 7, respectively). The
radial velocity is in units of km s$^{-1}$. The MVC were extracted from the
line profiles for the H$\alpha$, [O I]$\lambda$6300, [S II]$\lambda$6716
and [N II]$\lambda$6548 lines (whose symbols are on the top part of
the diagram; we note that, in this diagram, [O I]=[O I]$\lambda$6300,
[S II]=[S II]$\lambda$6716 and [N II]=[N II]$\lambda$6548). See the
text for a discussion.}
\label{radial}
\end{figure*}

Two distinct evolutionary times were chosen in order to investigate the
influence of the presence of an internal working surface (IWS) close
to the region where the the slits were placed. In particular, at $t=10$
years (see Fig. \ref{emaps}), an IWS is just being formed (close to the
jet inlet, in regions I-IV) due to the chosen $\tau_{pul}=8$ years.
On the other hand, for $t=15$ years, the regions I-IV are spatially
detached from the position of the nearest IWS (see Fig. \ref{slits}).
In Figure \ref{radial} we show the radial velocity of the MVC for regions
I, II, III and IV (from top to bottom) as a function of distance from
the jet axis (from $-0\farcs2$ to $0\farcs2$; the 0 in the abscissa
corresponding to the position of the jet axis) for $t=10$ years (left) and
$t=15$ years (right). The radial velocity of the MVC was extracted from
the profiles of the H$\alpha$, [O I]$\lambda$6300, [S II]$\lambda$6716
and [N II]$\lambda$6548 emission lines, using the procedure described
above. Although the MVC from different emission lines shows slightly
different behavior, the distribution of the radial velocity on both
sides of the jet axis for a given emission line is perfectly symmetric
for the M1 model, regardless the analyzed region or evolutionary time,
as we can see clearly in Figure \ref{radial}.

\begin{figure*}
\centering \hskip -4truecm
\includegraphics[width=17cm]{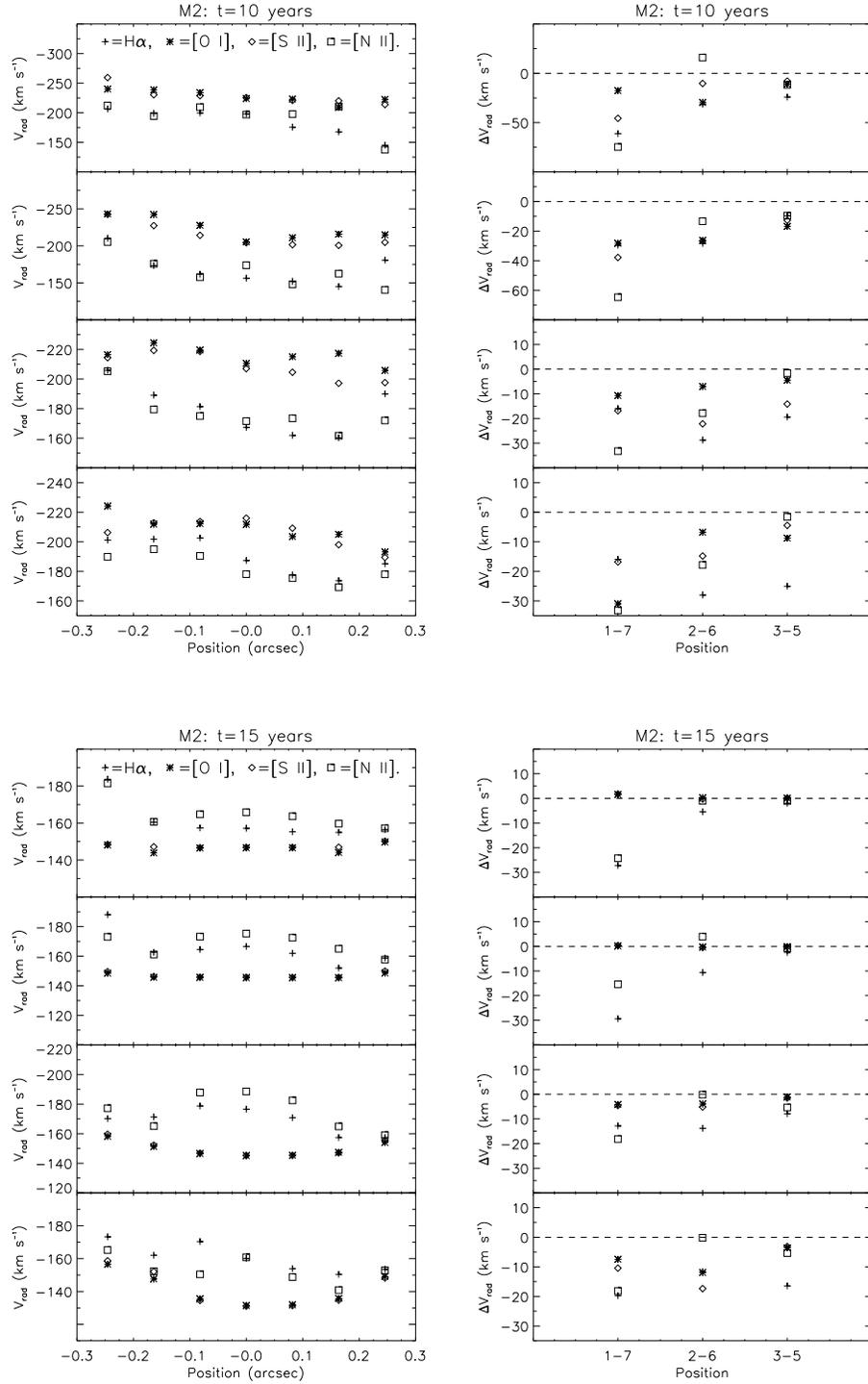}
\vskip -8.5truecm
\caption{{\it Left}) The same as in Figure \ref{radial}, but for the
M2 model (see Table \ref{tab1}) at $t=10$ years (top) and $t=15$
years (bottom).  The symbols and the scales are the same as in
Fig. \ref{radial}. {\it Right}) The radial velocity difference between
symmetrically disposed slits (with respect to the jet axis, slit 4;
see Figure \ref{slits}), for $t=10$ years (top) and $t=15$ years
(bottom). We note that $\Delta V_{rad}=V_{rad,i} - V_{rad,j}$, where
$i=1,2,3$ and $j=7,6,5$ define the slits according to the definition
in Figure \ref{slits}.}
\label{radial1}
\end{figure*}

In Figure \ref{radial1} we show the radial velocity ($V_{rad}$; left)
for regions I, II, III and IV (from top to bottom) of model M2 (with
precession but with no rotation) as a function of distance from the
jet axis (from $-0\farcs3$ to $0\farcs3$; as in the previous 
diagram\footnote{We note that the precession introduces a small shift on
the jet axis regarding the central pixel in the x-direction in our
simulations, i.e., x=65. However, due to the very small precession
angle chosen (see Table \ref{tab1}), the displacement of the slits with
respect to the jet axis is of the order of $\lesssim 0\farcs02$. This
small value should introduce negligible errors in the radial velocity
determinations. Moreover, this value is quantitatively comparable with
the deviation from the jet axis in observations of Bacciotti et al;
see Figure 8 in Bacciotti et al. (\cite{bacci02}).}), as well as the
difference ($\Delta V_{rad}$) between the radial velocity taken at
symmetrical positions with respect to the jet axis (right), at $t=10$
years (top diagrams) and $t=15$ years (bottom diagrams). The precession
of the jet axis introduces an asymmetry in the radial velocities as
a function of distance from the jet axis, as we can see in the left
diagrams of Figure \ref{radial1}. This effect is more pronounced at
$t=10$ years, as indicated by the difference in the radial velocities,
$\Delta V_{rad}$, in Fig. \ref{radial1} (top right diagram), and this
is due to the presence of an IWS close to regions I-IV. However, such an
asymmetry in the distribution of the radial velocity is still present at
$t=15$ years. As we can see in Figure \ref{radial1} (bottom diagrams),
$\Delta V_{rad}$ is systematically negative for most of the computed
emission lines. Although we cannot establish an unique behaviour for all
of the computed lines (considering the distinct regions and evolutionary
times), we note that some of them show  a clear trend of smaller $\Delta
V_{rad}$ for smaller distances from the jet axis ($V_{rad,3}-V_{rad,5}
< V_{rad,2}-V_{rad,6} < V_{rad,1}-V_{rad,7}$).

\begin{figure*}
\centering \hskip -4truecm
\includegraphics[width=17cm]{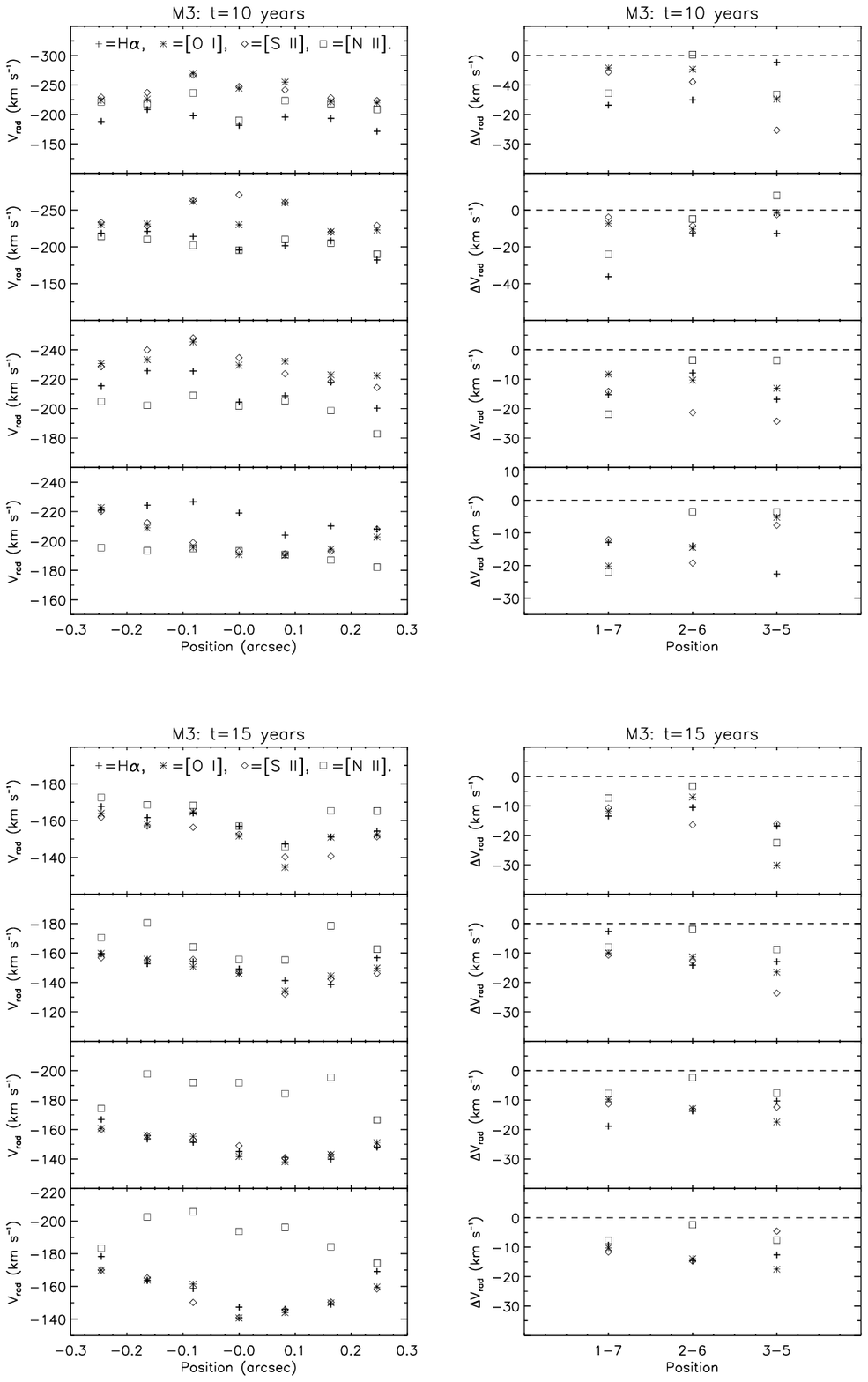}
\vskip -8truecm
\caption{The same as in Figure \ref{radial1}, but for the M3 model
(see Table \ref{tab1}).}
\label{radial2}
\end{figure*}

In Figure \ref{radial2} we again show the radial velocity ($V_{rad}$;
left) as a function of distance from the jet axis for regions I, II, III
and IV (from top to bottom), as well as the difference ($\Delta V_{rad}$)
between the radial velocity taken at symmetrical positions with respect
to the jet axis (right), at $t=10$ years (top diagrams) and $t=15$ years
(bottom diagrams), but now for model M3.  This model presents variability
in the velocity of injection as well as an initially imposed rotational
profile inside the jet beam, but has no precession (see \S 2 and Table
\ref{tab1}). The asymmetry in the radial velocity distribution, with
respect to the jet axis, can be clearly seen in Fig. \ref{radial2}. In
particular, $\Delta V_{rad}$ is negative for almost all emission
lines, regions and evolutionary stages (with the exception of the [N
II]$\lambda$6548 emission line, at $t=10$ years in region II).  However,
in opposition to model M2, $\Delta V_{rad}$ appears to be greater for
smaller distances from the jet axis (at least for some emission lines).

\begin{figure*}
\centering \hskip -4truecm
\includegraphics[width=17cm]{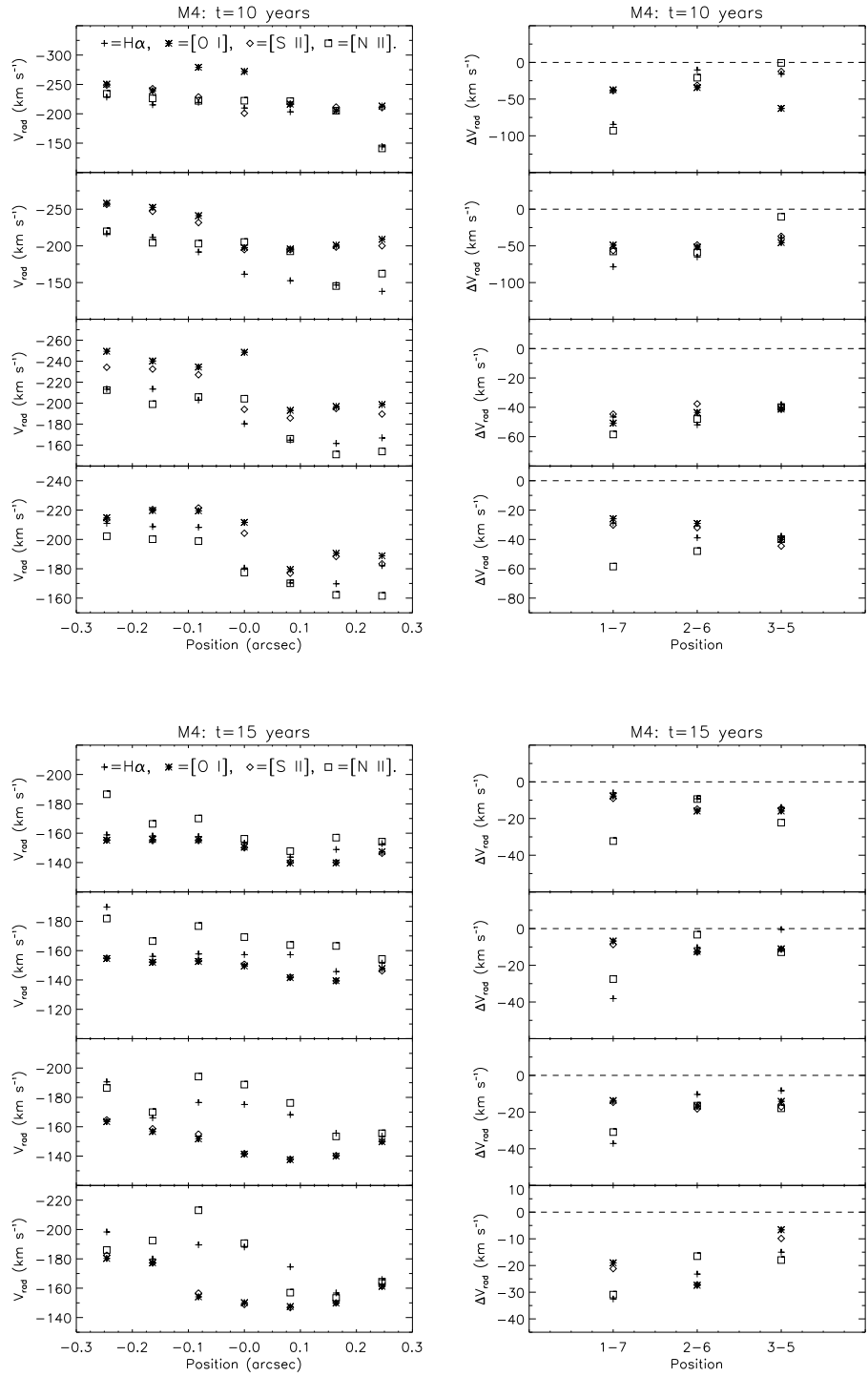}
\vskip -8truecm
\caption{The same as in Figure \ref{radial1}, but for the M4 model
(see Table \ref{tab1}).}
\label{radial3}
\end{figure*}

We present the radial velocity distribution for model M4 in Figure
\ref{radial3}. Model M4 has variability in both direction and velocity
of injection, as well as an initially imposed rotation profile inside
the jet beam (see Table \ref{tab1}).  The asymmetry in the radial
velocity distribution on both sides of the jet axis is more pronounced
than in models M2 and M3. This is clearly seen from the radial velocity
difference diagrams (right panels in Figure \ref{radial3}), which show
$\Delta V_{rad}$ of the order of 0 up to $\approx$ -80 km s$^{-1}$.
As in model M2, there is a general trend in the $\Delta V_{rad}$ data
that shows that it increases as a function of distance from the jet axis,
although such a trend is not followed by the H$\alpha$, [O I]$\lambda$6300
and [S II]$\lambda$6716 emission lines in region IV at $t=10$ years and
region I at $t=15$ years (see the right diagrams of Figure \ref{radial3}).

\begin{figure*}
\centering \hskip -3truecm 
\includegraphics[width=16cm]{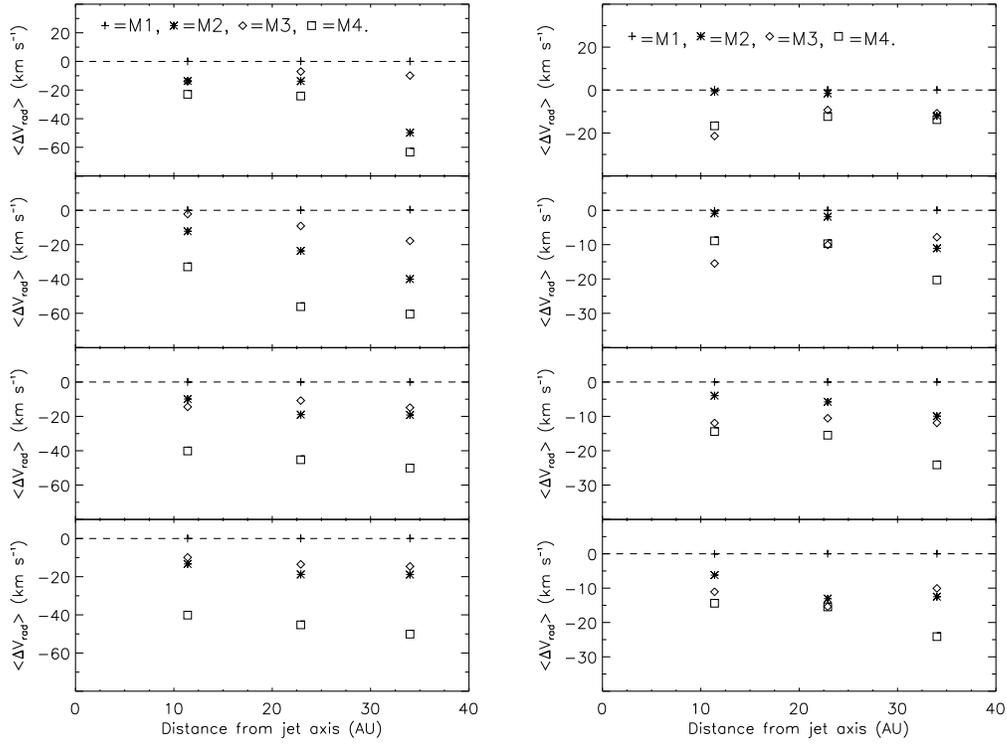}
\vskip -12truecm
\caption{The radial shift as a function of the distance from
the jet axis for regions I, II, III and IV (from top to
bottom), for models M1 (crosses), M2 (asterisks), M3 (diamonds) and
M4 (squares), for $t=10$ years (left) and $t=15$ years (right).}
\label{radial4}
\end{figure*}

In Figure \ref{radial4} we show the radial velocity shift as a function of
distance from the jet axis for regions I to IV (from top to bottom), and
for models M1 (crosses), M2 (asterisks), M3 (diamonds) and M4 (squares),
for $t=10$ years (left) and $t=15$ years (right). Each point in this
diagram corresponds to a mean value for the radial velocity shifts,
assuming that $\Delta V_{rad}= \sum_{l=1}^4 \Delta V_{rad,l}/4$, where
the different values of $l$ indicate the H$\alpha$, [O I]$\lambda$6300,
[N II]$\lambda$6548 and [S II]$\lambda$6716 emission lines. We clearly
see a general trend of increasing radial velocity shifts for larger
distances from the jet axis, followed by all the models (with the
expected exception of model M1, which shows a null $\Delta V$, as
discussed previously). Such a trend is similar to the one discussed
in Pesenti et al. (\cite{pesenti}) for their disk solution model
(and also followed in part by the data of Bacciotti et al \cite{bacci02}).

\section{Discussion and conclusions}

We have presented a set of 3D numerical simulations of variable velocity
jets, from which we have obtained detailed predictions of the emission
line profiles of a set of emission lines (H$\alpha$, [O I]$\lambda$6300,
[S II]$\lambda$6716 and [N II]$\lambda$6548). The models include an
axisymmetric jet, a precessing jet, a rotating jet, and, finally a jet
with both precession and rotation (see Table 1 and section 2).

Following Bacciotti et al. (\cite{bacci02}), we carry out double-Gaussian
fits to the line profiles, and then study the behaviour of the radial
velocity of the MVC (medium velocity component) as a function of distance
from the jet axis, for several positions along the jet flow.  We find
that while the axisymmetric model of course predicts radial velocities
whith a symmetric behaviour on both sides of the jet axis, the other
three models predict asymmetric behaviours.

The models with precession and with rotation predict systematic offsets of
$\vert \Delta V_{rad} \vert \lesssim 20$~km~s$^{-1}$ for the MVC between
symmetric positions on both sides of the jet axis.  These systematic
offsets are seen at most positions along the jet axis, and for many
of the emission lines. The magnitudes of these offsets are similar to
the ones observed (for many positions and emission lines) by Bacciotti
et al. (\cite{bacci02}), Coffey et al. (\cite{coffey04}) and Woitas et
al. (\cite{woitas05}). In the model with both precession and rotation
(in which the precession was assumed to have a prograde direction), the
two effects combine to give larger offsets of $\sim 40-80$~km~s$^{-1}$.

We have also computed a model with rotation and a retrograde
precession. This model (not presented in this paper) produces side-to-side
offsets in the MVC radial velocity which are much smaller, and which do
not show clear trends as a function of distance from the jet axis.

From these results, we conclude that the radial velocity asymmetries
observed in the DG Tau, TH28, RW Aur and LkH$\alpha$321 microjets (see
Bacciotti et al.  \cite{bacci02}; Coffey et al. \cite{coffey04}; Woitas
et al. \cite{woitas05}) can be modeled either as due to a rotation of
the gas in the initial jet cross section, or as due to a precession
of the ejection direction. Interestingly, though a precession results
in a point-symmetry in the ``precession spiral'' morphology of the jet
beam, the direction of the resulting rotation of the ejected material
is of course the same one for both the jet and the counterjet. Also, a
``precession spiral'' morphology can be produced by an orbital motion
of the source around a binary companion (Masciadri \& Raga \cite{masb};
Gonz\'alez \& Raga \cite{gon04c}), resulting in similar radial velocity
patterns, but with a morphology with mirror symmetry between the jet
and the counterjet.

It is therefore unclear at this time whether the systematic radial
velocity asymmetries observed by Bacciotti et al. (\cite{bacci02}),
Coffey et al. (\cite{coffey04}) and Woitas et al. (\cite{woitas05})
in the cross sections of several microjets are due to jet rotation or
precession (or to a combination of both effects). In order to proceed
further, it will be necessary to proceed with more detailed modelling,
in which not only the radial velocity asymmetries are considered. A
study which also considers the locci of the jet and counterjet on the
plane of the sky, resulting in models which simultaneously reproduce the
morphology and the radial velocity structure of the observed outflows,
might be able to solve this problem. Of the four microjets observed by
Bacciotti et al. (\cite{bacci02}), Coffey et al. (\cite{coffey04}) and
Woitas et al. (\cite{woitas05}), the jet from RW Aurigae shows the chain
of knots with the best alignment. Because of this lack of evidence for a
precession spiral structure in the jet beam, the RW Aur jet is probably
the most certain case for an observed rotation of the jet beam.

Interestingly, the referee has pointed out to us the very recent results
of Cabrit et al. (\cite{cabrit05}), who detect a disk around RW~Aur
rotating in the opposite sense of the ``jet rotation'' seen by Woitas
et al. (\cite{woitas05}). This result in principle eliminates a jet
rotation interpretation for the side-to-side radial velocity asymmetries
observed in the RW~Aur jet. Therefore, the remaining possibilities are
that the jet is precessing (producing the asymmetries explored above),
or that it has another intrinsic side-to-side asymmetry (see, e.~g.,
Soker 2005 and Cabrit et al. 2005).

\begin{acknowledgements}

We would like to thank the anonymous referee for relevant and helpful
comments and suggestions. The work of A.H.C. is partially supported by a
CAPES fellowship (BEX 0285/05-6). A.H.C. and M.J.V. would like to thank
the PROPP-UESC (project 220.1300.327), PRODOC-UFBa (projects 991042-88
and 991042-108) and CNPq (projects 62.0053/01-1-PADCT III/Mil\^enio,
470185/2003-1 and 306843/2004-8) for partial financial support, as well
as the ICN-UNAM staff for hospitality. The authors acknowledge support
from CONACyT grants 41320-E, 43103-F and 46828 and the DGAPA-UNAM grant
IN113605. We thank Israel D\'\i az for maintaining our Linux server at
ICN-UNAM, where all the numerical simulations have been carried out.

\end{acknowledgements}

\vfill
\eject
\null

\vfill
\eject
\null

\vfill
\eject
\null

\vfill
\eject
\null

\vfill
\eject
\null

\vfill
\eject
\null

\vfill
\eject
\null

\vfill
\eject
\null

\end{document}